\documentclass[a4paper,fleqn,usenatbib]{mnras}

\usepackage{mathptmx}

\usepackage[T1]{fontenc}
\usepackage{ae,aecompl}

\usepackage{graphicx}	
\usepackage{savesym}
\usepackage{amsmath}	
\savesymbol{iint}
\savesymbol{iiint}
\savesymbol{iiiint}
\savesymbol{idotsint}
\usepackage{txfonts}
\restoresymbol{TXF}{iint}
\restoresymbol{TXF}{iint}
\restoresymbol{TXF}{iiint}
\restoresymbol{TXF}{iiiint}
\restoresymbol{TXF}{idotsint}

\usepackage{pdflscape}
\usepackage{epstopdf}
\usepackage{amssymb}	

\title[Characterization of 2020~AV$_\mathbf{2}$]{Physical characterization of 2020~AV$_\mathbf{2}$, the first known asteroid orbiting inside Venus orbit}
\author[M. Popescu et al.]{
M. Popescu$^{1,2}$\thanks{E-mail: popescu.marcel1983@gmail.com},
J. de Le\'on$^{2,3}$,
C. de la Fuente Marcos$^{4}$,
O. Vaduvescu$^{5,2}$,
\newauthor
R. de la Fuente Marcos$^{6}$,
J. Licandro$^{2,3}$,
V. Pinter$^{5,7}$,
E. Tatsumi$^{8,2}$,
O. Zamora$^{2,3}$,
\newauthor
C. Fari{\~n}a$^{5,2}$,
and L. Curelaru$^{9}$
\\
$^{1}$ Astronomical Institute of the Romanian Academy, 5 Cu\c{t}itul de Argint, 040557 Bucharest, Romania\\
$^{2}$ Instituto de Astrof\'{\i}sica de Canarias (IAC), C/V\'{\i}a L\'{a}ctea s/n, 38205 La Laguna, Tenerife, Spain\\
$^{3}$ Departamento de Astrof\'{\i}sica, Universidad de La Laguna, 38206 La Laguna, Tenerife, Spain\\
$^{4}$ Universidad Complutense de Madrid, Ciudad Universitaria, E-28040 Madrid, Spain\\
$^{5}$ Isaac Newton Group of Telescopes (ING), Apto. 321, E-38700 Santa Cruz de la Palma, Canary Islands, Spain\\
$^{6}$ AEGORA Research Group, Facultad de CC. Matem\'{a}ticas, Universidad Complutense de Madrid, Spain\\
$^{7}$ School of Doctoral Sciences, The University of Craiova, Str. A. I. Cuza nr. 13, 200585 Craiova, Romania\\
$^{8}$ Department of Earth and Planetary Science, The University of Tokyo, Bunkyo-ku, Tokyo, Japan\\
$^{9}$ Amateur Astronomer, Bra\c{s}ov, Romania
}

\date{Accepted XXX. Received YYY; in original form ZZZ}

\pubyear{2020}

\begin{document}
\label{firstpage}
\pagerange{\pageref{firstpage}--\pageref{lastpage}}
\maketitle

\begin{abstract}
The first known asteroid with the orbit inside that of Venus is 2020~AV$_{2}$. This may be the largest member of a new population of small bodies with the aphelion smaller than 0.718~au, called Vatiras. The surface of 2020~AV$_{2}$ is being constantly modified by the high temperature, by the strong solar wind irradiation that characterizes the innermost region of the Solar system, and by high-energy micrometeorite impacts. The study of its physical properties represents an extreme test-case for the science of near-Earth asteroids. Here, we report spectroscopic observations of 2020~AV$_{2}$ in the 0.5-1.5~$\mu m$ wavelength interval. These were performed with the Nordic Optical Telescope and the William Herschel Telescope. Based on the obtained spectra, we classify 2020~AV$_{2}$ as a Sa-type asteroid.  We estimate the diameter of this Vatira to be $1.50_{-0.65}^{+1.10}$ km by considering the average albedo of A-type and S-complex asteroids ($p_V=0.23_{-0.08}^{+0.11}$), and the absolute magnitude (H=$16.40\pm0.78$ mag). The wide spectral band around 1~$\mu m$ shows the signature of an olivine rich composition. The estimated band centre $BIC = 1.08 \pm 0.02~\mu m$ corresponds to a ferroan olivine mineralogy similar to that of brachinite meteorites.
\end{abstract}




\section{Introduction}

Solar system numerical simulations predict the existence of a population of asteroids that have their orbits entirely within that of Venus \citep{2012Icar..217..355G}. These objects with the aphelion in the interval (0.307, 0.718)~au are called Vatiras (to distinguish them from the Atira-class near-Earth objects) or Interior to Venus Orbit Objects. According to the NEOSSat-1.0 orbital distribution model \citep{2012Icar..217..355G}, they could represent about 0.22 per cent of the steady-state near-Earth asteroids (NEAs), and they have a high probability ($\approx$80 per cent) of originating in the $\upsilon_6$ secular resonance with Saturn, in the main asteroid belt.

Dynamical pathways that enable NEA transitions from one orbital realm into the other have been reported by \cite{2019RNAAS...3..106D, 2019MNRAS.487.2742D, 2020MNRAS.494L...6D}. They found that most predicted Vatiras and Atiras are former Atens (Earth-crossing asteroids with semi-major axis smaller than 1~au and aphelion larger than 0.983~au). These pathways are exemplified by those of asteroids such as 2019~AQ$_3$, 2019~LF$_6$, and 2018~JB$_3$. However, the authors could not discard that some fraction of Vatiras may have remained in their current locations since the time when the inner Solar system became stable. These objects may be subjected to a particular type of secular resonance, the von Zeipel-Lidov-Kozai oscillation \citep{1910AN....183..345V, 1962P&SS....9..719L, 1962AJ.....67..591K}  that make them long-term stable.

%
%
     \begin{table}
      \centering
      \fontsize{8}{11pt}\selectfont
      \tabcolsep 0.15truecm
      \caption{Heliocentric Keplerian orbital elements of 2020~AV$_{2}$ and their 1$\sigma$ uncertainties. The orbit determination 
               is referred to epoch JD 2459000.5 (2020-May-31.0) TDB (Barycentric Dynamical Time, J2000.0 ecliptic and equinox). 
               Source: JPL Small-Body Database (solution date, 2020-May-21 11:41:22).
              }
      \begin{tabular}{ccccc}
       \hline
        Orbital parameter                                 &   & value$\pm$1$\sigma$ uncertainty \\
       \hline
        Semimajor axis, $a$ (au)                          & = &   0.55542$\pm$0.00010           \\
        Eccentricity, $e$                                 & = &   0.1771$\pm$0.0003             \\
        Inclination, $i$ (\degr)                          & = &  15.872$\pm$0.008               \\
        Longitude of the ascending node, $\Omega$ (\degr) & = &   6.707$\pm$0.004               \\
        Argument of perihelion, $\omega$ (\degr)          & = & 187.31$\pm$0.02                 \\
        Mean anomaly, $M$ (\degr)                         & = & 222.49$\pm$0.09                 \\
        Perihelion, $q$ (au)                              & = &   0.4571$\pm$0.0002             \\
        Aphelion, $Q$ (au)                                & = &   0.65377$\pm$0.00012           \\
        Absolute magnitude, $H$ (mag)                     & = &  16.40$\pm$0.78                   \\
       \hline
      \end{tabular}
      \label{elements}
     \end{table}
%
%

The only known Vatira, 2020~AV$_{2}$, was first observed by the Zwicky Transient Facility on January 4, 2020 \citep{2020MPEC....A...99B}. The preliminary orbit of this asteroid is described by the aphelion at 0.654~au, the perihelion at 0.457~au, and an orbital inclination of 15.9~$\degr$ (according to JPL Small-Body Database, accessed on May 21, 2020). Currently (as of May 21, 2020), no other properties are determined for this object, except those derived from astrometric measurements which cover an arc of 19 days. These values are summarized in Table~\ref{elements}. 

The dynamical history of this object has been explored using $N$-body simulations \citep{2020MNRAS.494L...6D}. The results show that 2020~AV$_{2}$ was a former Atira-class, and perhaps a former Aten-class asteroid, which reached the Vatira orbit relatively recently in astronomical terms, $\sim10^5$ yr (within $9\sigma$ confidence level). Its aphelion distance has been decreasing over the last 1 Myr, after experiencing some close encounters with Venus and Mercury. Now, it is detached from direct interaction with Venus and it is subjected to secular interactions with the Earth-Moon system and Jupiter.  However, they also reported that it is statistically possible that 2020~AV$_{2}$ could be currently subjected to the 3:2 mean-motion resonance with Venus, or more likely just outside this resonance. Similar results have also been reported by \citet{2020MNRAS.493L.129G}.

The numerical simulations discussed by \citet{2020MNRAS.494L...6D} indicate that the main direct perturber of  2020~AV$_{2}$  is Mercury and it may have experienced close encounters under 0.007~au with it. The orbital evolution is quite regular, but not fully stable due to the encounters with Mercury. Around the time of capture into the Vatira orbital realm, 2020~AV$_{2}$ experienced multiple distant --- beyond the Hill radius of the planets --- encounters with both Mercury and Venus. Such encounters led to its present orbital state. The orbit of this body makes it subjected to high temperature, strong solar wind irradiation characteristic for the innermost region of the Solar system, and energetic micrometeorites bombardment. All these processes are constantly modifying the surface of 2020~AV$_{2}$, in a way only comparable to some extent with the one experienced by Mercury's surface \citep[e.g.][]{2019JGRE..124.2326B}.  Thus, the study of the physical properties of 2020~AV$_{2}$ represents a unique opportunity to investigate an extreme test-case for the science of NEAs and it will provide much needed information on the conditions of the space-environment at 0.55~au.

The total number of discovered asteroids having the orbit confined within Earth's orbit (i.e. with the aphelion smaller than the Earth's perihelion of 0.983~au) is 22. The largest ones (considering the absolute magnitude) are 
(163693) Atira (H = 16.3 mag), the representative member of this population, and (418265) 2008~EA$_{32}$ which has H = 16.4 mag. The rest of them span an absolute magnitude range from 17.3 to 21.2 mag (by considering an albedo of $p_V=0.15$, this interval translates into an equivalent diameter range of 0.2 to 1.2 km). For the small sizes, this statistics is biased in favour of those with the best visibility windows (i.e. highest apparent brightness and largest values of the solar elongation). The absolute magnitude of 2020~AV$_{2}$, H = $16.40~\pm~0.78$ mag (reported by JPL Small-Body Database Browser, accessed on May 21, 2020) indicates an object with a size larger than one kilometer. The bright absolute magnitude of this object has favoured its discovery, but this value is one of the brightest compared with other members of the Atira-class (within the uncertainty, it may be the brightest), or Aten-class.

The predicted number of Vatiras is 0.2 - 0.3 per cent of the steady state NEAs population \citep{2012Icar..217..355G, 2018Icar..312..181G}.  There are 979 discovered NEA (according to MPC website accessed on March 25, 2020) with H$\leq$17.9~mag, and 546 NEAs with H$\leq$17.1~mag -- which is the faintest magnitude limit for 2020~AV$_{2}$. These values are identical to the number of $962_{-56}^{+52}$ predicted by \citet{2018Icar..312..181G}. Based on these statistics and estimates, 2020~AV$_{2}$ is the largest Vatira that may exist and just another two or three bodies with H in the range of 17 to 18 magnitudes should exist. 

In this context, we started an observational program with the aim of determining the observable properties of 2020~AV$_{2}$. These observations were performed with the 2.56~m Nordic Optical Telescope (NOT) and the 4.2~m William Herschel Telescope (WHT), both located at El Roque de los Muchachos Observatory (ORM) in La Palma, Canary Islands (Spain). Carrying out these observations was very challenging due to the low maximum elongation of this target (when observed, the solar elongation of 2020~AV$_{2}$ was close to 37~$\degr$). Here, we present these new and unique results regarding the physical properties of 2020~AV$_{2}$. This article is organized as follows. To put in context these findings, we present in Section 2 the latest update for the dynamical evolution which we computed following the same approach as in \citet{2020MNRAS.494L...6D}. Section 3 describes the data acquisition and reduction procedures. Section 4 analyzes the observed spectra and the photometry.  The implications of our findings are discussed in Section 5. Section 6 summarizes our conclusions.

%
\section{Updated dynamic properties of 2020~AV$_{2}$}
\begin{figure}
       \centering
        \includegraphics[width=\linewidth]{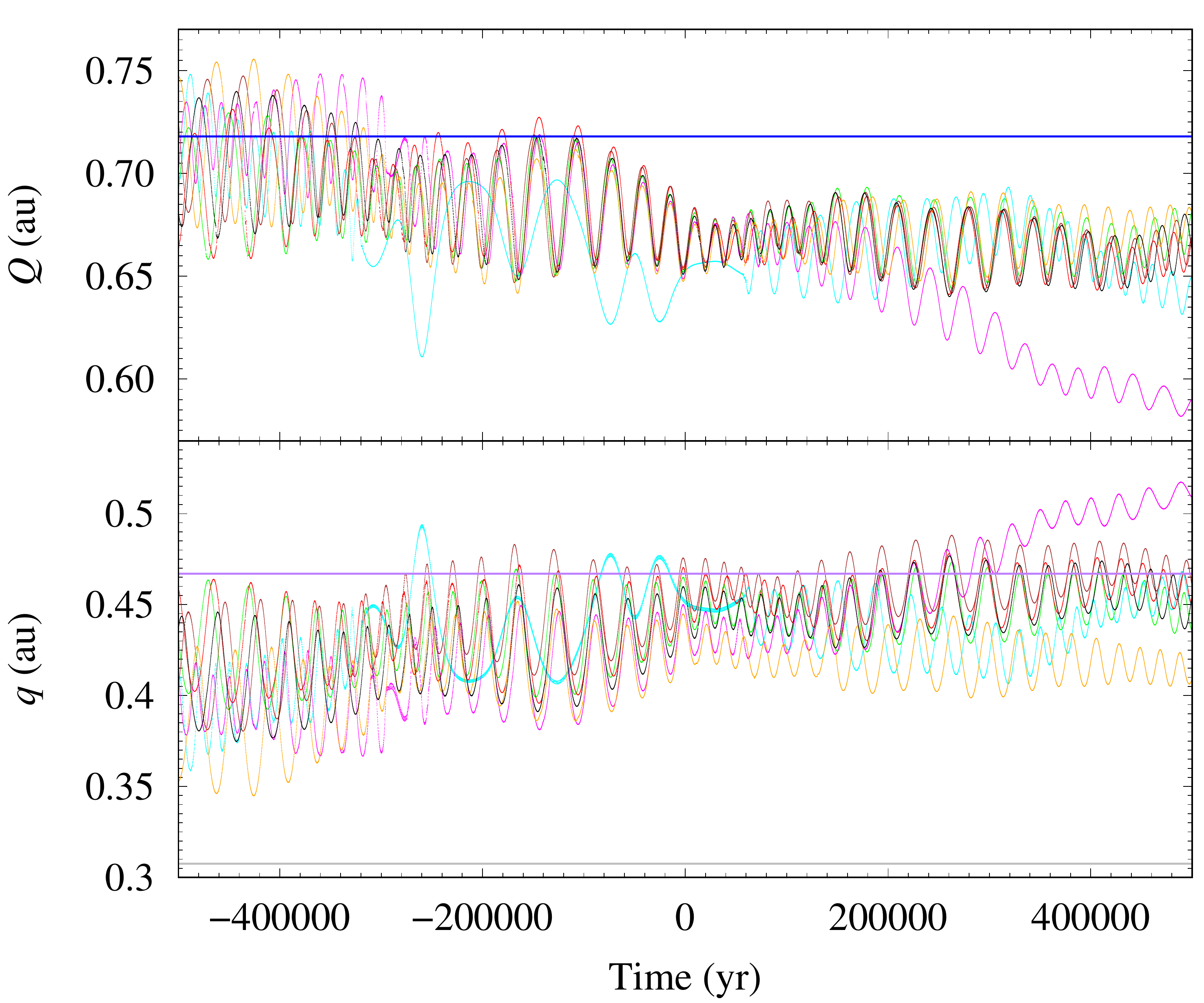}
        \caption{Evolution of the aphelion ($Q$, top panel) and perihelion ($q$, bottom panel) distances of the nominal orbit (in 
                 black) of 2020~AV$_{2}$ as described by the orbit determination in Table~\ref{elements} and those of control 
                 orbits with Cartesian vectors separated $+$3$\sigma$ (in green), $-$3$\sigma$ (in cyan), $+$6$\sigma$ (in red), 
                 $-$6$\sigma$ (in magenta), $+$9$\sigma$ (in brown), and $-$9$\sigma$ (in orange) from the nominal values in 
                 Table~\ref{vector}. In blue (top panel), we show the perihelion distance of Venus, and in purple and grey (bottom 
                 panel), we indicate the aphelion and perihelion distances of Mercury.
                }
        \label{Qq}
     \end{figure}

\begin{figure}
       \centering
        \includegraphics[width=\linewidth]{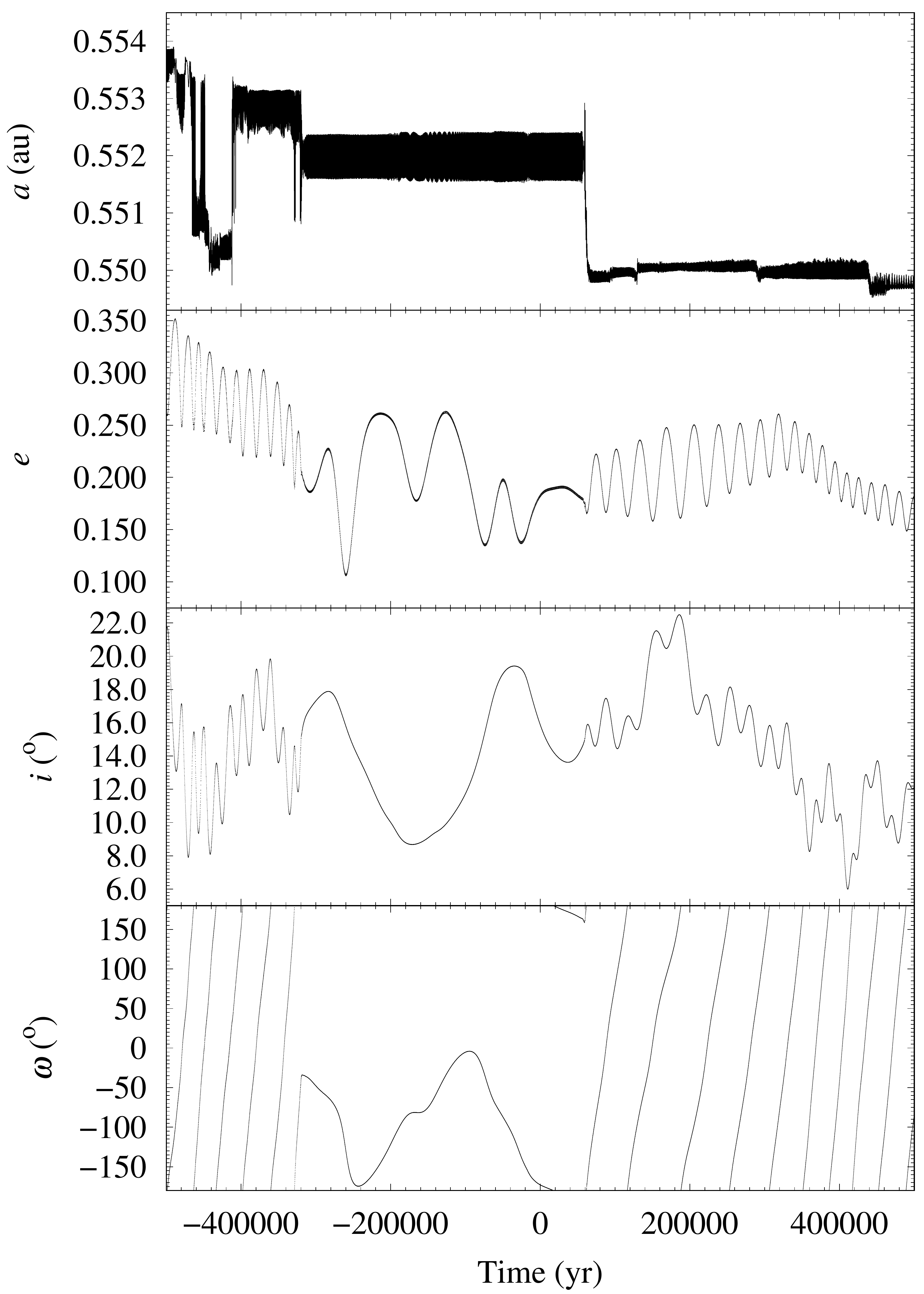}
        \caption{Evolution of the values of the semimajor axis ($a$, top panel), eccentricity ($e$, second to top panel), inclination ($i$,
                 second to bottom panel), and argument of perihelion ($\omega$, bottom panel) for a relevant control orbit of 2020~AV$_{2}$,
                 one with Cartesian vector components $-$3$\sigma$ away from the nominal one in Table~\ref{vector}. The object remains 
                 trapped inside the 3:2 mean-motion resonance with Venus at 0.552~au for nearly 400~kyr. The values have been computed with 
                 respect to the invariable plane of the system (see \citealt{2020MNRAS.494L...6D} for additional details).
                }
        \label{resonance}
     \end{figure}

%
    The orbit determination of 2020~AV$_{2}$ has recently been slightly improved (see Table~\ref{elements}) with respect to the 
    one used by \citet{2020MNRAS.494L...6D} ---although the data-arc span remains at 19~d, the number of observations increased 
    from 135 to 139. Using data provided by Jet Propulsion Laboratory's Solar System Dynamics Group Small-Body Database (JPL's 
    SSDG SBDB, \citealt{2015IAUGA..2256293G})\footnote{\url{https://ssd.jpl.nasa.gov/sbdb.cgi}}, we have repeated some of the 
    calculations described in \citet{2020MNRAS.494L...6D} and confirmed earlier results. Figure~\ref{Qq} shows that as pointed out 
    by \citet{2020MNRAS.493L.129G} and \citet{2020MNRAS.494L...6D}, 2020~AV$_{2}$ is a statistically robust present-day Vatira 
    because $Q$$<$0.718~au. The present-day 2020~AV$_{2}$ can only experience close encounters with Mercury as it became detached 
    from Venus about 10$^{5}$~yr ago (Fig.~\ref{Qq}, top panel). The new orbit determination does not place 2020~AV$_{2}$ inside 
    the 3:2 mean-motion resonance with Venus at 0.552~au, at least nominally. However, the control orbit with Cartesian vector 
    components $-$3$\sigma$ away from the nominal one (in cyan in Fig.~\ref{Qq}) is trapped in this resonance (see 
    Fig.~\ref{resonance}, top panel). Therefore and although improbable, it cannot be discarded that 2020~AV$_{2}$ is subjected to 
    such a resonant motion. In any case, this object has remained inside the Atira region for an extended period of time, being 
    affected by physical conditions like average temperature and solar irradiance levels second only to those experienced by 
    Mercury among known Solar system objects. 

%
%
     \begin{table}
      \centering
      \fontsize{8}{11pt}\selectfont
      \tabcolsep 0.15truecm
      \caption{Cartesian state vector of 2020~AV$_{2}$ used in the calculations with components and 1$\sigma$ uncertainties. Epoch 
               as in Table~\ref{elements}. Source: JPL's SBDB.
              }
      \begin{tabular}{ccccc}
       \hline
        Component                         &   &    value$\pm$1$\sigma$ uncertainty                                \\
       \hline
        $X$ (au)                          & = &    4.468136099398687$\times10^{-1}$$\pm$5.31406753$\times10^{-4}$ \\
        $Y$ (au)                          & = &    4.394341120331902$\times10^{-1}$$\pm$4.05282352$\times10^{-4}$ \\
        $Z$ (au)                          & = &    1.071580973001556$\times10^{-1}$$\pm$1.83925659$\times10^{-4}$ \\
        $V_X$ (au~d$^{-1}$)                      & = & $-$1.546825399980614$\times10^{-2}$$\pm$1.78745447$\times10^{-5}$ \\
        $V_Y$ (au~d$^{-1}$)                      & = &    1.206086274388701$\times10^{-2}$$\pm$2.14978061$\times10^{-5}$ \\
        $V_Z$ (au~d$^{-1}$)                      & = &    3.920518738301038$\times10^{-3}$$\pm$3.52818858$\times10^{-6}$ \\
       \hline
      \end{tabular}
      \label{vector}
     \end{table}
%
%
%
%
%
%
%
%
%

\section{Observations and data reduction}

The main objective of our observational program was to obtain spectroscopic data in the visible and near-infrared (NIR) region for characterizing the composition of 2020~AV$_{2}$. The log of observations is shown in Table~\ref{obslog}. Auxiliary images using the J and Ks filters were obtained during the evenings of January 13 and 14. These exposures were used to constrain some photometric properties, summarized in Table~\ref{obslogPhotom}.

This asteroid was observed at an altitude lower than 25~$\degr$ above the horizon, and the data acquisition started during the astronomical twilight, on the evenings of January 11, 13, and 14. We emphasize that a better observing geometry cannot be achieved using ground-based telescopes. The next favourable visibility window for this object will  be on August 2021, when it will reach a maximum solar elongation of  $\approx39.9\deg$ (according to Minor Planet Center website, accessed on May 21, 2020).

\begin{table*}
\centering
\caption{Log of spectroscopic observations. The table includes the instrument used, the observing date, the UT at the start of observation, the exposure time, the airmass for the first and the last acquired image (Start-Am, End-Am), the apparent $V$ magnitude, the phase-angle ($\alpha$), and the asteroid heliocentric distance ($r_{\sun}$). The values for $V-mag$, $\alpha$, $r_{\sun}$ were retrieved from the Minor Planet Ephemeris Service.
}
\label{obslog}
\begin{tabular}{l l  l l c c c c c}
\hline
Instrument & Obs. date  & Start-UT & Exp-time[s] & Start-Am & End-Am & V-mag & $\alpha[\deg]$ & $r_{\sun}[au]$\\ \hline
NOT/ALFOSC & 2020-01-11 & 19:19 & 2x900 & 2.3 & 2.7 & 18.0 & 94.4 & 0.605 \\
WHT/ACAM   & 2020-01-11 & 19:35 & 1x900 + 1x511 & 2.7 & 4.6 & 18.0 & 94.4 & 0.605 \\
WHT/LIRIS  & 2020-01-13 & 19:29 & 19 x 120 & 2.3 & 4.9 & 18.0 & 96.9 & 0.598 \\
\hline
\hline
\end{tabular}    
\end{table*}

\begin{table}
\centering
\caption{Observing log for the LIRIS/WHT photometric images. The observing date, the filter, the total number of images (N), the exposure time ($t_{exp}$), and the full width at half maximum for the asteroid ($FW_{a}$) and for the nearby stars ($FW_{s}$) are shown.}
\label{obslogPhotom}
\begin{tabular}{l c c c c c}
\hline
Obs. Date & Filter  & N & $t_{exp}[s]$ & $FW_{a}[pix]$ & $FW_{s}[pix]$ \\ \hline
Jan. 13 & $Ks$ & 13 & 10;15 & 4.6 & 4.76$\pm$0.27 \\
Jan. 13 & $J$ & 8 & 15 & 6.3 & 6.59$\pm$0.31 \\
Jan. 14 & $Ks$ & 8 & 15;20 & 3.9 & 4.07$\pm$0.19 \\
Jan. 14 & $J$ & 19 & 20 & 5.2 & 5.45$\pm$0.29 \\
\hline
\hline
\end{tabular}    
\end{table}

\subsection{Spectroscopic observations}

The ALFOSC (Alhambra Faint Object Spectrograph and Camera) instrument mounted on the NOT was used to obtain a visible spectrum. The spectrograph was used with the Grism\_\#4 and with a slit having a width of 1.8 arcsec. The second order blocking filter GG475 was added to avoid second order contamination. This setup covers the optical interval 0.48 - 0.92~$\mu m$ with a resolution  of $\sim$ 300.  The calibration images (biases, internal flats, and arcs for lamps with He, Ne, and ArTh) were obtained at the end of the night.  The  solar analogue stars HD 218676, HD 206559, and SA 115-271 were observed at a similar airmass as the asteroid, in order to correct for the telluric absorption bands and to obtain the asteroid reflectance spectrum.

Another visible spectrum was also obtained using the ACAM (Auxiliary-port CAMera) instrument which is mounted permanently at a folded-Cassegrain focus of the WHT. The second order-blocking filter GG495A was used to obtain the uncontaminated spectrum over 0.495 - 0.95~$\mu m$. The 400-lines/mm transmission VPH (Volume Phase Holographic) grating and a slit width of 1.5 arcsec provide a resolution of $\sim$ 400. The calibration images were obtained before and after the target observations. The arcs were taken with the telescope at target's position to avoid wavelength shifts due to flexure. The solar analog stars (HD 218676, SA 115-271, HD 212809) used for calibration were observed at similar airmass as the asteroid.

The LIRIS (Long-slit Intermediate Resolution Infrared Spectrograph) instrument was used to obtain the NIR spectrum. This is a NIR imager/spectrograph mounted at the Cassegrain focus of the WHT. It has a 1k$\times$1k HAWAII detector covering the 0.8 to 2.5~$\mu m$ wavelength range.  For our purpose, we used the $lr\_zj$ grism and a slit-width of 2.5 arcsec. This configuration allows to obtain spectra over 0.9 to 1.5~$\mu m$ spectral region with a resolution of $\sim$700. The spectra were obtained alternatively at two separate locations along the slit denoted A and B, following
the nodding procedure. The images were acquired with short exposure times (120~s for the asteroid and 5-10~s for the calibration stars). The same solar analogues used for visible observations (HD 218676, SA 115-271, HD 212809) were observed for computing the asteroid reflectance spectrum. The second attempt, made on the night of January 14, allowed only to obtain photometric exposures using $J$ and $Ks$ filters, before the telescope was closed due to high humidity. 

The data reduction followed the standard procedures for both the visible and NIR exposures. Preprocessing of the CCD images consisted in bias and flat-field correction (the flat fields were obtained using the internal lamps of each telescope/instrument, which are dedicated to this purpose). The consecutive NIR images, taken in the A and B position, were subtracted (A-B, B-A) to remove the sky. The GNU Octave software package \citep{octave} was used to create scripts for IRAF -- Image Reduction and Analysis Facility \citep{1986SPIE..627..733T} to perform all these tasks~automatically. The extraction of the 1D-spectrum from the images was made with the \emph{ IRAF - apall} package. Each spectral exposure was inspected to avoid any artefact such as the contamination due to background stars, target tracking errors, or spurious reflections.

\begin{figure*}
\includegraphics[width=\textwidth]{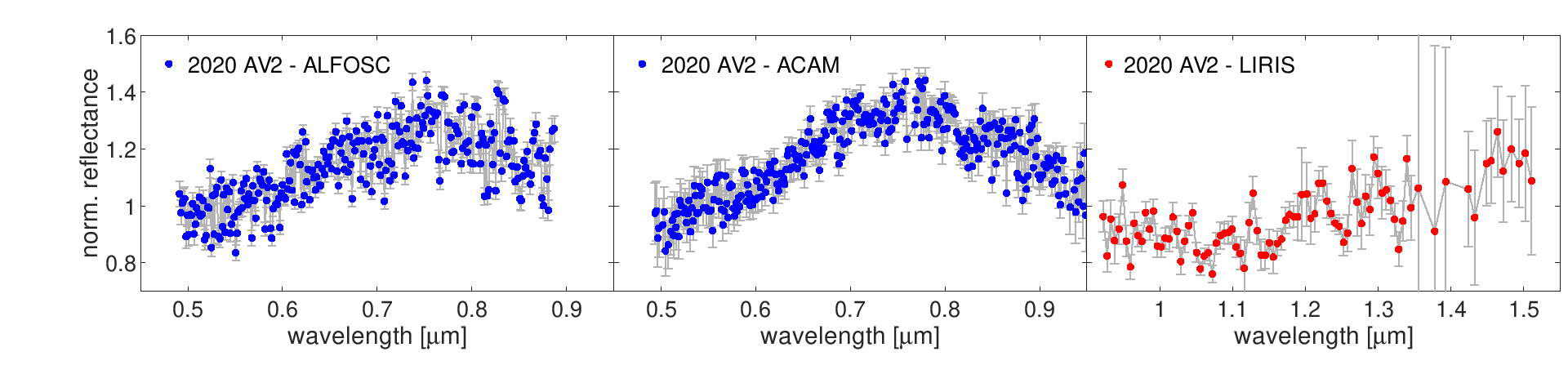}
\caption{The observed spectral data of 2020~AV$_{2}$. The left panel shows the data obtained with the ALFOSC/NOT instrument, the middle panel shows the data obtained with the ACAM/WHT instrument, and the right panel shows the ones obtained with the LIRIS/WHT. The visible spectra are normalized at 0.55~$\mu m$, while the NIR one at 1.25~$\mu m$.}
 \label{specplot}
\end{figure*}

The wavelength map was determined using the emission lines from the lamps available for each set of observations. These lamps contain  Helium, Neon, and Thorium-Argon for ALFOSC/NOT\footnote{\url{http://www.not.iac.es/instruments/alfosc/lamps/}}, 
Copper-Argon and Copper-Neon for ACAM/WHT\footnote{\url{http://www.ing.iac.es/Astronomy/instruments/acam/observing.html}}, and Xenon and Argon for LIRIS/WHT\footnote{\url{http://www.ing.iac.es/Astronomy/instruments/liris/calibrations.html}}. 

The last step was to obtain the reflectance spectrum. This was done by dividing the observed asteroid spectrum by the spectra of the three solar analogue stars. The possible wavelength shifts between the asteroid and the solar analogue (the so called "heartbeats") were corrected by an additional routine. All solar analogues provided similar results in terms of spectral shape. This validates the observations and the data reduction process (we note that SA 115-271 is a solar analogue used frequently for calibrating asteroid reflectance spectra). 

The final spectrum was obtained by averaging the reflectance spectra obtained from each exposure. A binning procedure of four pixels for the visible spectra and of nine pixels for the NIR ones was applied to increase the signal to noise ratio (SNR). We removed part of the measurements close to the  spectral interval edges for the ALFOSC/NOT data due to large error bars. The results are shown in Fig.~\ref{specplot}. The visible reflectances were normalized to unity at 0.55~$\mu m$, and those containing the NIR part were normalized at 1.25~$\mu m$.

\subsection{Photometric data}

Four groups of images (a group comprises the images obtained with a single filter during an observing night) were obtained in the $J$ and $Ks$ bands by using the WHT/LIRIS instrument (Table~\ref{obslogPhotom}). These~auxiliary exposures were acquired for target identification and centring in the slit during the spectral observations. The field of view of LIRIS is 4.27~arcmin $\times$ 4.27~arcmin and the pixel scale is 0.25~arcsec/pixel.

LIRIS images were taken following a dithering pattern. The pre-processing of the raw images consisted in flat field corrections. The consecutive images (which are designated A and B and they have a positional dither), obtained with the same filter, were subtracted one from the other to remove the sky background. After applying this procedure, the remnant sky patterns were removed using the \emph{astnoisechisel} procedure (from GNU Astronomy package, version 0.11) with the default parameters \citep{gnuastro}. The \emph{astnoisechisel} is a noise-based non-parametric technique dedicated to the detection of very faint nebulous objects which are buried in noise. This tool detects the sky background as an intermediary step, and allows to remove it.

\begin{figure}
\includegraphics[width=\columnwidth]{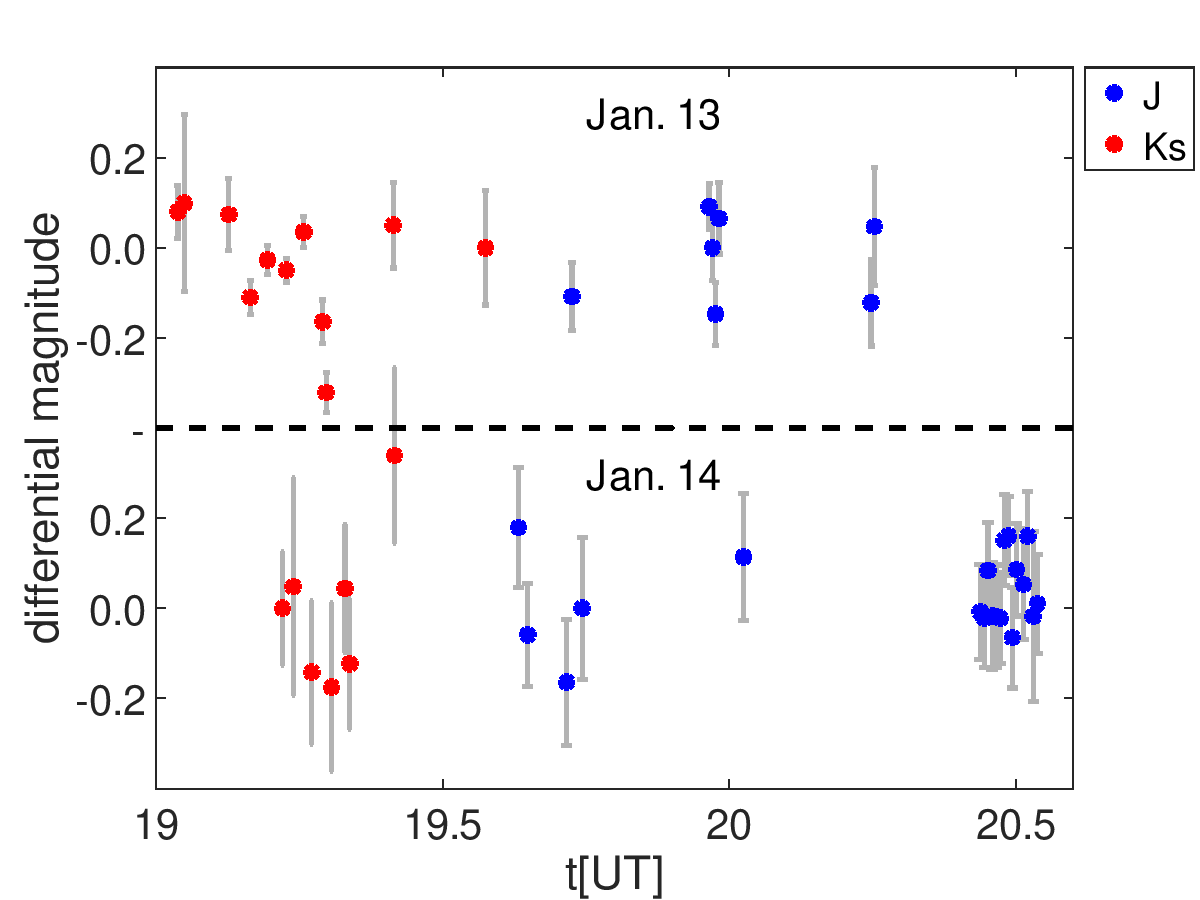}
\caption{The differential magnitudes of 2020~AV$_{2}$ derived from the images acquired with LIRIS. Two filters were used (the observation obtained with $J$ are shown in blue, while those with $Ks$ are shown in red). Each set of data (observations performed with a single filter on a night) is shifted to zero (median value of all the values in a set is shifted to zero).}
 \label{photomplot}
\end{figure}

The differential and absolute photometry was performed using the IRAF, \emph{apphot} package \citep{1986SPIE..627..733T}. The differential photometry technique consisted in obtaining measurements of the asteroid and of 5-7 reference stars. For this task we used an aperture of 6-10 pixels (fixed for each set of images). Then, the magnitude differences were calculated (the constant offset is subtracted) for finding luminosity variations. These differential magnitudes are shown in Fig.~\ref{photomplot}. The data only show a lightcurve variation within $\pm$ 0.2~mag over each 30 to 60~min group of points. We note that we removed four outlier (one from each set) observations due to a difference larger than $\sim$0.4~mag compared to the neighbouring points (such discontinuities are not expected for an asteroid lightcurve, and are likely observational artefact).   

The calibration to the absolute magnitudes was done using the 2MASS All-Sky Catalog of Point Sources. There were 6-7 sources considered for calibrating each image, in order to compute the $(J-Ks)$ colour. However, the results obtained for the two nights are not consistent.

\section{Results}

The spectral and the photometric observations of 2020~AV$_{2}$ were investigated using various techniques. We present the taxonomic classification, the mineralogical analysis, and the comparison with the Relab database. The results are used to constrain the size of the body, its surface temperature, and its composition.

The analysis of the spectral data has to take into account the influence of the high phase angle. To discuss this effect on various spectral features, we need to start our analysis with a rough approximation for the taxonomic type and based on the findings, we can arrive to a final classification. Because the observation geometry was an unusual one, we introduced in this section the comparison of different models and results reported in the literature for the phase angle effects.

\subsection{Spectral properties}

The reflectance spectra of 2020~AV$_{2}$ in the visible and NIR (up to 1.5 $\mu m$) interval are shown in Fig.~\ref{specplot}. The spectrum in the visible wavelengths was obtained with two different instruments, ALFOSC/NOT and ACAM/WHT. The differences between the two observations are lower than the noise level. The uncertainties associated to observational procedures may be caused by the instrument setup (centring in the slit, alignment of the slit, telescope tracking accuracy), by the data reduction procedures (imperfect flat-field correction, trace identification, sky subtraction), or the selection of solar analogues used for obtaining the relative reflectances. In this case, the similarity of these two observations (performed with the target at low altitudes) provides reliability to our findings and limits the systematic errors. 

\begin{figure*}
\includegraphics[width=\textwidth]{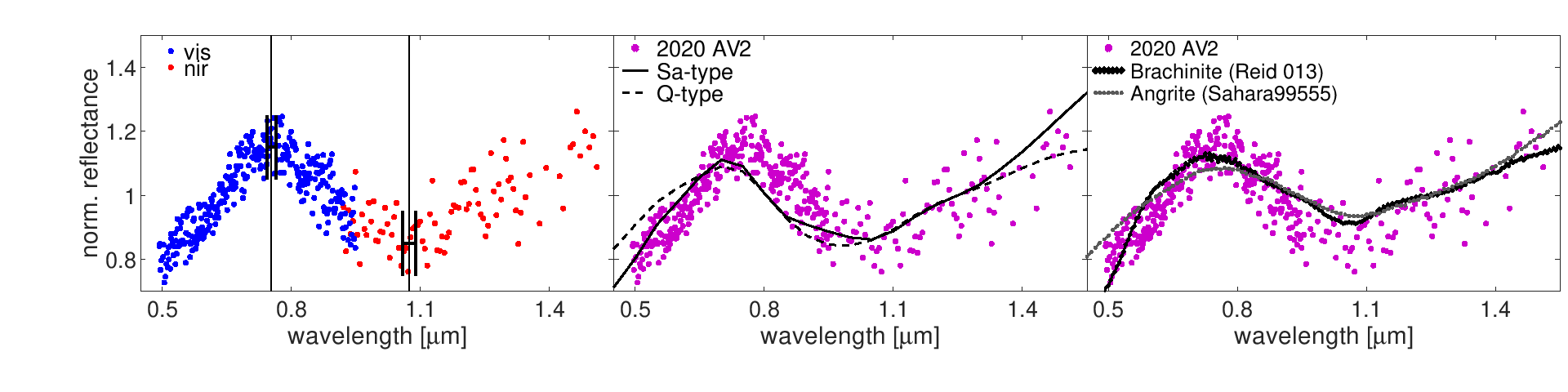}
\caption{Spectral matching of 2020~AV$_{2}$. All spectra are normalized at 1.25~$\mu m$. The combined visible - near infrared spectrum is shown on the left-hand side panel. The positions of the reflectance maximum at 0.745~$\mu m$ and of the band minimum at 1.075~$\mu m$, together with the associated error-bars, are shown by the two markers. The comparison with the Sa-type and the Q-type is shown in the middle panel. The best meteorite spectral fit is shown on the right-hand panel.}
 \label{taxonm}
\end{figure*}

The analysis presented here uses the visible spectral data obtained with ACAM/WHT because it has lower error-bars and it covers a longer wavelength interval (0.49 - 0.95)~$\mu m$. This spectrum was merged with the NIR data obtained with LIRIS/WHT. The merging procedure consisted in multiplying the NIR spectrum with a coefficient such that the median reflectance values, computed over the common wavelength interval (0.92 - 0.95)~$\mu m$, are identical for both spectra. The composite spectrum is shown in Fig.~\ref{taxonm} -- left panel, and it was normalized to unity at 1.25~$\mu m$.

At first glance, the NIR band shown by the composite spectrum points toward an olivine-pyroxene composition and indicates a taxonomic classification within the S-complex (S, Sa, Sq, Sr, Sv types) or end members A, O, Q, R, V. The solution to this conundrum is determined below by taking into account the large phase angle and the position of the spectral features.

The first step in analyzing the spectrum is to compute the visible spectral slope ($BR_{slope}$) over the 0.5 - 0.7~$\mu m$ spectral interval (this interval avoids the beginning of the NIR $\sim1\mu m$ band). The "red" value found, $BR_{slope}~=~17.7\pm0.8\%/0.1 ~\mu m$ is intermediate between the slope of S-complex asteroids $BR^S_{slope}~=~12.2\pm3.3\%/0.1 ~\mu m$ and that of A-types $BR^A_{slope}~=~20.9\pm2.3\%/0.1 ~\mu m$. We calculated these average slopes for the taxonomic classes by using iteratively the same procedure for the 371 asteroid spectra (out of which six objects are classified as A-types and 199 are classified as S-complex) reported by \cite{2009Icar..202..160D}. The uncertainty is one standard deviation of the values computed for the reference spectra of a given class. We note that the large majority of the 371 reference spectra are main belt asteroids observed at phase angles between 10~$\degr$ and 40~$\degr$.

This $BR_{slope}$ provides a way to quantify the reddening effects that can be explained by phase angle, space-weathering and grain size. The information available for this object does not allow to disentangle these effects. The phase reddening effect on olivine-pyroxene compositions was studied by \citet{2012Icar..220...36S} using the slope over the (0.45 - 2.5)~$\mu m$ spectral interval. They found that ordinary chondrite meteorites are the most affected by the phase reddening, displaying a large variation of the slope. This effect is also outlined in the data brought by the Hayabusa space mission for the S-type asteroid (25143) Itokawa \citep{2018Icar..311..175T}. In this case, the spectral slope varied by a factor of two, in a non-linear manner, for phase angles between 5~$\degr$ to 40~$\degr$.

The phase angle effects over the visible to NIR spectral interval were discussed by \citet{2019Icar..324...41B} using the NEAs spectra obtained during the MIT-Hawaii Near-Earth Object Spectroscopic Survey. For the individual objects, they found strong correlation (a correlation coefficient of 0.94) between the spectral slope and phase angle which confirms the phase reddening effect.  However, the correction coefficients depend on the object and a relation applicable to all asteroids could not be found. The same conclusion, that there is no general relation for the phase reddening effect on asteroids, is reached for the NEA's spectra reported by other works \citep[e.g.][]{2010A&A...517A..23D,2019A&A...627A.124P}. 

Thus, we cannot quantify the phase angle effect for 2020~AV$_{2}$, we only confirm that it exists and it is in the range reported by other studies. However, an approach similar to ours, based on $BR_{slope}$,  was performed by \citet{2018P&SS..157...82P} using the spectra of NEAs smaller than 300~m. The slope of 2020~AV$_{2}$ is in the range of values they found for the S-complex objects observed at $\sim90\deg$ phase angle.

The second step of our analysis searched the position of the spectral features. These are the maximum of reflectance, and the minimum of the absorption band (denoted as $BImin$). A smoothed spectral curve is required for their computation. The smoothing is obtained by applying the \emph{splinefit} function from GNU Octave package. The procedure was repeated 1000 times by randomizing each point from the spectrum within its error. For each of these runs, the two extrema are computed. We report the median value and the standard deviation for the position of the maximum and of the minimum.

Using the technique described above, we found the band minimum at $BImin~=~1.075\pm0.015~\mu m$ (Fig.~\ref{taxonm}). The $BImin$ represents a rough approximation for the $1~\mu m$ band centre ($BIC$), which is one of the main spectral features used to diagnose the surface composition of asteroids \citep{1986JGR....9111641C, 1993Icar..106..573G,2010Icar..208..789D}. In order to compute the $BIC$, the continuum must be removed from the spectral curve. This is determined by a linear fitting between the two maxima enclosing the band. Thus, the $BIC$ is the minimum of the spectral curve after dividing it by this linear continuum. The first maximum was found to be  at $0.745\pm0.01~\mu m$. However, the second maximum cannot be computed due to the limited wavelength coverage of the 2020~AV$_{2}$ spectrum. To obtain a proxy estimate for the second maximum we linearly extrapolated the spectral trend for wavelengths longer than 1.1~$\mu m$. Then, we computed the continuum between the first maximum and the reflectance extrapolated value at 1.7~$\mu m$. This wavelength approximation for the second reflectance maximum, took into account the spectra presented by \citet{2014Icar..228..288S}, which shows a peak at $1.68\pm0.06~\mu m$. After the spectral continuum removal, we estimated $BIC~=~1.080~\mu m$ for 2020~AV$_{2}$.

For evaluating the uncertainty introduced by this method a Monte-Carlo approach was considered. We repeated the procedure by considering the value of the second maximum as different extrapolated reflectances corresponding to a wavelength position in the range of 1.55 - 1.80~$\mu m$. This gives an estimation uncertainty of $\sigma_{BIC}=0.020~\mu m$.

The value determined for the $BIC = 1.08 \pm 0.02$ reveals a composition similar to the S(I) subtype of asteroids with olivine-pyroxene mixtures, defined by \cite{1993Icar..106..573G}.  These objects exhibit a strong $1~\mu m$ absorption feature, with a nonexistent or negligible 2~$\mu m$ feature. Furthermore, with a confidence level of 2  $\sigma_{BIC}$,  this feature is in the range of values shown by the olivine rich/dominated asteroids \citep{2014Icar..228..288S}. 

The spectral band parameters are influenced by temperature and phase angle effects. \citet{2012Icar..220...36S} reported that the variations of band centres with increasing phase angle are smaller than the uncertainties associated with their test data, thus for our case this correction can be disregarded.

An estimation for the subsolar equilibrium surface temperature is given by Eq.~\ref{temp} \citep{2009M&PS...44.1331B}. 
\begin{equation}
T = [\frac{(1-A)L_0}{16\eta\epsilon\sigma\pi r^2}]^{1/4}
\label{temp}
\end{equation}
where $L_0$ is the solar luminosity (3.827$\times10^{26}$ W), $\eta$ is the beaming parameter (assumed to be 1), $\epsilon$ is the asteroid's infrared emissivity ($\approx$0.9), $\sigma$ is the Stefan-Boltzman constant (5.67$\times10^{-8}\frac{J}{sm^2K^4}$), and $r$ is the heliocentric distance (shown in Table~\ref{obslog}). Assuming an average albedo $A$ = 0.2, typical for asteroids with olivine-pyroxene composition, the temperature $T=\sim350$ K is found. This temperature implies a wavelength correction  $\Delta BIC=0.003~\mu m$  for the band centre, following the equations presented by \cite{2014Icar..228..288S}. This correction is one order of magnitude less than our uncertainty $\sigma_{BIC}$ of $0.020 \mu m$, and can be safely ignored.

The spectral parameters outlined above allow us to conclude that 2020~AV$_{2}$ is an olivine-rich asteroid. Future observations in the 2~$\mu m$ region will determine if the composition is fully dominated by olivine or a small fraction of pyroxene may exist.

To test the inferring of an olivine rich asteroid, we compared the spectrum of 2020~AV$_{2}$ with the entire Relab\footnote{\url{http://www.planetary.brown.edu/relab/}} data base of meteorites \citep{2004LPI....35.1720P, 2016LPI....47.2058M}. The first 10 best spectral matches are with brachinite, angrite, and pallasite meteorites. The first two of these matches are shown in Fig.~\ref{taxonm} - right panel. They correspond to the Reid 013 (sample ID: MT-TXH-052-A, chip particulated $\leq$ 45~$\mu m$) brachinite meteorite, and to Sahara 99555 (Sample ID: TB-TJM-057, particulated $\leq$ 125~$\mu m$) angrite meteorite. The brachinites are olivine achondrites, distinguishable by their olivine composition $Fo_{65-70}$, coupled with the absence or very low abundance of Ca-poor pyroxene (ranging from 0 to 10 per cent in volume). They formed from a chondritic precursor, by different depletions in metal-sulfide and basaltic components. The angrites are silica under-saturated calcium-rich basaltic achondrites with a granular textured olivine pyroxenite made of Al/Ti diopside, Ca-bearing olivine and Fe-Al spinel. The pallasites are stony iron meteorites with olivine in the range $Fo_{82-70}$, texturally variable and heterogeneous. Polished surfaces of these meteorites show a metallic mesh enclosing cm-sized olivines. These statements about meteorites are based on \citet{2004mete.book.....H} and references there in.

The solutions obtained by comparison with laboratory spectra of meteorites are not unique due to limited spectral range. Also, they ignore the temperature and the phase angle effects. Nevertheless, all of these matches constrain the composition of 2020~AV$_{2}$ to an olivine rich asteroid with an achondritic nature coming from a differentiated body with igneous formation history. 

The last step of the spectral analysis is to assign a taxonomic classification. This is done by performing a curve fitting with all the 24 (25 - with the recent updates) classes corresponding to Bus-DeMeo taxonomy \citep{2009Icar..202..160D}. The best matches correspond to Sa and Q spectral types.  

The spectrum of 2020~AV$_{2}$ shows a very red spectral $BR_{slope}$, which seems incompatible with the Q-type, but this can be interpreted  as being caused by phase reddening effect (as already discussed above). The Q-type is common in the NEA population \citep[e.g.][]{2019Icar..324...41B}. Moreover, \citet{2019A&A...627A.124P} have shown that they are predominant on orbits with perihelia close to the Sun. This spectral type represents fresh surfaces \citep{2010Natur.463..331B}, and a composition similar to the LL ordinary chondrite meteorites. This type of composition would be expected if it has an origin in the Flora family \citep{2008Natur.454..858V}. The new laboratory results of \citet{2019PASJ...71..103H} show that Q-type asteroids may not necessarily represent a fresh surface, but large (more than 100~$\mu m$) ordinary chondritic particles with space weathering exhibit spectra consistent with this taxonomic class. Nevertheless, the position of $BImin~=~1.075\pm0.015~\mu m$ discards the Q-type classification (which would have been expected considering the statistics of NEAs population) and a possible association with ordinary chondrites. From the work done by \cite{2012Icar..220...36S} with spectra of LL chondrites at different phase angles, they conclude that the whole VNIR spectra gets redder with increasing angle, and this translates into a shift to shorter wavelengths of the minimum of the 1~$\mu m$ band. So, this implies that, if corrected from phase reddening, the absorption band minimum of our spectrum will move to even longer wavelengths.

The 1~$\mu m$ band shape is most compatible with a Sa-type. This is a class with a deep and extremely broad absorption 1~$\mu m$ band and similar features as A-types but it is less red \citep{2009Icar..202..160D}. This spectral type is very rare and it is indicative of a very high olivine content \citep[e.g.][]{2019Icar..322...13D, 2018MNRAS.477.2786P}, which is in agreement with our previous analysis.Thus, we can conclude that the most likely classification is a Sa-type.

There are very few asteroids classified as Sa-type with their albedo determined. \citet{2010AJ....140..933R} and \citet{2011ApJ...741...90M} list only one object with $p_V$ in the range 0.339 to 0.397 (depending on the model). This value may represent an outlier, considering that Sa is an intermediate class between S-type and A-types, which have the albedo of $p_V^S$ = 0.211 $\pm$ 0.068, and $p_V^A$ = 0.191 $\pm$ 0.034 respectively \citep{2011ApJ...741...90M}. A reasonable statistics (39 asteroids) for determining an average $p_V$ = 0.230 $\pm$ 0.099 is available for the objects classified as Sa in Bus taxonomy \citep{2002Icar..158..146B}, although this classification is based only on visible spectra.

The diameter of 2020~AV$_{2}$ can be estimated using the expression \citep[e.g.][]{2007Icar..190..250P}, $D = 1329\cdot(p_V)^{-0.5}\cdot10^{-0.2H}$. A value of 1.5 km is obtained by considering H = 16.4~mag and $p_V$ = 0.230. The large uncertainty in the absolute magnitude ($\sigma_H = 0.78$) and of albedo $p_V~\in$ (0.15 -0.34), translates into an interval of possible values for the equivalent diameter $D~\in$ (0.85, 2.6) km.

The albedo value allows an estimation for the range of subsolar equilibrium surface temperature (Eq.~\ref{temp}). This shows an asteroid roasted at temperatures in the range of 330 $\pm$ 10 K (at aphelion Q = 0.654~au) to 395 $\pm$ 15 K at perihelion (q = 0.454~au).

\subsection{Photometric properties}

The combined images, using track and stack method, for the night of January 14, 2020 show the average apparent magnitudes $J_{mag}$ =  17.129$\pm$0.091 mag, and $Ks_{mag}$ =  16.136$\pm$0.074 mag. The colours can be computed  using the closest in time images for avoiding lightcurve variations. We obtained $(J-Ks)_{n2}$ = 1.01$\pm$0.23, where the two exposures with $J$ and $Ks$ filters were taken at an interval of $\Delta t$ = 12.99~min. This colour is in agreement with the classification as an olivine-rich object, when compared with the average $(J-Ks)$ value of A- and Sa- type asteroids \citep{2018A&A...617A..12P}.

However, the colour measurement based on the first night photometric exposures shows inconsistencies. The same method gave a $(J-Ks)_{n1}$ = 0.16$\pm$0.15~mag, where the two observations were made during an interval of $\Delta t$ = 9.06~min. A likely explanation for the discrepancy between $(J-Ks)_{n1}$ and $(J-Ks)_{n2}$ is the variable atmosphere during the observations, while the dispersion of the photometric points  (Fig.~\ref{photomplot}) of this night of January 13 indicate unaccounted errors. Although less likely, we cannot exclude that these differences may be caused by real lightcurve variations or peculiar reflections due to the large phase angle of the observations.

We note that the ephemerides computed using MPC website, predict an apparent visual magnitude $V_{mag}$ = 18.0 mag for both nights. For the first night, the difference between the predicted $V_{mag}$ and the observed average $J^{n1}_{mag}$ = 16.14~mag is 1.86 mag. This value is larger than the expected range for the asteroids, and suggest that the $(J-Ks)_{n1}$ is doubtful. However, the difference between the predicted $V_{mag}$ and the observed average $J^{n2}_{mag}$ for the night of January 14 is 0.87 mag which is typical for asteroids, as outlined by large surveys data \citep{2016A&A...591A.115P}.  This gives confidence for the $(J-Ks)_{n2}$ = 1.01$\pm$0.23 determination.

In order to detect a possible cometary activity, we measured the full width at half maximum (FWHM) on the combined images. The results are listed in Table~\ref{obslogPhotom}. In these images, the asteroid presented a point source profile (the FWHM of the stars is slightly larger due to the fact that the telescope was in differential tracking mode, following the asteroid).  Furthermore, in order to search for very faint cometary activity we applied the \emph{astnoisechisel} procedure. The results confirmed the stellar-like profile (point source).

\section{Discussion}

The 1.5 km size of 2020~AV$_{2}$ is uncommon when considering that bodies orbiting close to the Sun may experience catastrophic disruption over time because of the high temperature. \citet{2016Natur.530..303G} discussed several possible mechanisms that can be responsible for the reduced number of NEAs having a low perihelion orbit. One possibility is that rocks break into small grains by thermal cracking \citep{2014Natur.508..233D} and the resulting dust is removed by radiation pressure. Another possibility is that the anisotropic emission of thermal photons makes the asteroid to spin faster, phenomenon known as the Yarkovsky--O'Keefe--Radzievskii--Paddack (YORP) mechanism  \citep[e.g.][]{2015aste.book..509V}, to the point when gravity and cohesive forces can no longer keep it intact. However, the images obtained during our observations show a stellar profile (see Section 4.2). The low SNR of our photometric data does not allow to detect any photometric trend. However, the composition does not exclude a binary object. For example,  (136617) 1994 CC is a triple NEA with a diameter of $\sim$650 m for the primary, $\sim$10 m for the secondary, and $\sim$5 m for the tertiary component, which shows an olivine dominated composition \citep{2011P&SS...59..772R}.

The olivine rich asteroids, such as  2020~AV$_{2}$, are rare in the NEA population. There are only six objects (two A-types and four Sa-types) compatible with this composition in the sample of 1040 NEOs reported by \citet{2019Icar..324...41B} for the MITHNEOS program. The largest olivine dominated asteroid which orbits close to the NEA population is (1951) Lick \citep{2004A&A...422L..59D} and it has a diameter of 5.6 km. 

These olivine asteroids are more abundant for sizes smaller than 300-m, where they can represent up to 5.4~per cent as reported by \citet{2018MNRAS.477.2786P}. This result was confirmed by \citet{2019AJ....158..196D} who reports a fraction of 3.8~per cent. Most of these olivine rich asteroids are expected to have been formed through magmatic differentiation \citep[e.g.][]{2007M&PS...42..155S,2014Icar..228..288S}. Thus, they were the major constituents of the mantles of most differentiated bodies, and their paucity in the observational data is known as the "Missing Mantle Problem" \citep[e.g.][]{2015aste.book...13D}. Our finding, of a $\sim$1~km-sized body with olivine rich composition, on a peculiar orbit, is in favour of the 'battered-to-bits' scenario \citep{1996M&PS...31..607B}. These authors proposed that the mantle and crustal material of the original differentiated bodies had been ground down to pieces below the limit of detectability of current spectroscopic surveys in the main asteroid belt.

The signature of olivine rich composition, the 1~$\mu m$ spectral shape, is a broad absorption feature formed by three individual bands. The deep central absorption band, which is clearly distinguishable in our spectrum, is located around 1.04~$\mu m$ and it is attributed to $Fe^{2+}$ transitions in the M2 crystal site. The other two weaker bands, should be located at about 0.85 and 1.3~$\mu m$ and corresponds to the  $Fe^{2+}$ transitions in the M1 crystal site \citep{1993macf.book.....B}. Due to the low SNR of our 2020~AV$_{2}$ spectrum, we are not able to identify these weaker features. \citet{1987JGR....9211457K} report a linear relation between the $BIC$ and $Fe^{2+}$ content. Our determined $BIC = 1.08 \pm 0.02$ is compatible with a ferroan olivine. \citet{2007M&PS...42..155S} have shown that ferroan olivine cannot be produced  from  the  melting  and  segregation  of  ordinary chondrites because the starting materials must already be FeO-rich. They note that even moderately ferroan olivine likely originate from oxidized R chondrite or melts thereof. Following \citet{2003M&PS...38.1601M}, an  alternative  history  for ferroan olivine asteroids and for their analogous brachinites meteorites is that an oxidation event occurred during melting and converted Fe-metal to FeO. Nebular processes can produce such a composition.

Numerical simulations show that 2020~AV$_{2}$ has been in the Vatira region during the last $\sim$ 1.5~kyr. According to JPL Small-Body Database Browser in the recent years there were frequent approaches to Mercury at distances of the order of 0.07~au. Much closer encounters may have happened in the more distant past.  Such flybys, although they took place at perihelion implying high relative velocities (in the range of 8 to 20 km/s), may trigger resurfacing events due to tidal forces \citep{2016Icar..268..340C}. The effect of these depends on the texture of the asteroid \citep{2006Icar..183..312S}.

The surface of 2020~AV$_{2}$ is also affected by high-energy micrometeorite impacts. \citet{2018ApJ...868...74S} have shown evidence for a circumsolar dust ring near Mercury's orbit. In this dynamical context, we may argue that on every orbit 2020~AV$_{2}$ crosses the ring and suffers the impact of myriads of dust grains that may freshen up the surface by kinematic erosion. The resulting dust is then removed by dissipative forces, such as radiation pressure and corpuscular drag. 

The Vatira 2020~AV$_{2}$ is a natural probe into this relatively unexplored region of the Solar system, and spectroscopic studies during favourable visibility windows may help in understanding how fast is the evolution of asteroid surfaces there and how it proceeds, physically.

\section{Conclusions}

In this paper, we have reported spectroscopic observations of the first known asteroid orbiting inside Venus orbit, namely 2020~AV$_{2}$. These observations were performed with the NOT and the WHT telescopes from El Roque de los Muchachos Observatory located in La Palma, Canary Islands (Spain). We merged the data obtained with different instruments in order to cover the 0.5 - 1.5~$\mu m$ wavelength interval.

The obtained spectrum shows a deep and broad absorption band around 1~$\mu m$, which is the signature of mafic minerals. Based on the obtained spectrum, we classified 2020~AV$_{2}$ as  Sa-type in the Bus-DeMeo taxonomic system. 

We used the average albedo of the A-type, and S-complex asteroids (according to the assigned class) to estimate the equivalent diameter as 1.5 km. The uncertainties of absolute magnitudes and albedos translate into a range of possible sizes between 0.85 to 2.6 km. The average albedo allows also to compute the range of subsolar equilibrium surface temperature of 2020~AV$_{2}$, from 330 $\pm$~10 K (at aphelion Q = 0.654~au) to 395 $\pm$ 15~K at perihelion (q = 0.454~au). 

The value estimated for the $BIC = 1.08 \pm 0.02~\mu m$ indicates an olivine-rich composition with ferroan mineralogy. This is confirmed by the spectral matching with the brachinite meteorites.  These findings reveal the achondritic nature of 2020~AV$_{2}$. Thus, it is fragment from a differentiated body with igneous formation history. 

The~auxiliary photometric data indicate a lightcurve variation of $\sim$0.4 mag within 30 to 60 min, and a point-like source. The $(J-Ks)$ = 1.01$\pm$0.23 colour obtained on the night of January 14, 2020 is in agreement with an olivine-rich object. 

\section*{Data availability}
The data underlying this article will be shared on reasonable request to the corresponding author. We intend to upload the data to a public repository accessible trough Virtual Observatory tools.
 
\section*{Acknowledgements}

This work was developed in the framework of EURONEAR collaboration and of ESA P3NEOI projects. The work of M.P. was supported by a grant of the Romanian National Authority for Scientific Research - UEFISCDI, project number PN-III-P1-1.2-PCCDI-2017-0371. M.P., J.dL. and J.L. acknowledge support from the AYA2015-67772-R (MINECO, Spain), and from the European Union’s Horizon 2020 research and innovation programme under grant agreement No 870403 (project NEOROCKS). We thank S. J. Aarseth for providing the code used in this research to study the dynamical evolution of 2020~AV$_{2}$ and A. I. G\'omez de Castro for providing access to computing facilities. This work was partially supported by the Spanish 'Ministerio de Econom\'{\i}a y Competitividad' (MINECO) under grant ESP2017-87813-R. Based on observations made with the William Herschel Telescope (operated by the Isaac Newton Group of Telescopes) and the Nordic Optical Telescope.  The LIRIS spectroscopy was obtained as part of SW2019b57 proposal. We are grateful to all the staff that helped us to carry out these difficult observations.
The paper make use of data published by the following web-sites Minor Planet Center\footnote{\url{https://www.minorplanetcenter.net/}}, JPL Small-Body Database Browser \footnote{\url{https://ssd.jpl.nasa.gov/sbdb.cgi}},  SMASS - Planetary Spectroscopy at MIT \footnote{\url{http://smass.mit.edu/}}, and Relab Spectral Database \footnote{\url{http://www.planetary.brown.edu/relab/}}. We thank the anonymous reviewer whose comments helped improve and clarify this manuscript.




\bibliographystyle{mnras}
\bibliography{2020AV2.bib} 



%


\bsp	
\label{lastpage}
\end{document}